# Percolation Modeling of Self-Damaging of Composite Materials

**Sergii Domanskyi** and **Vladimir Privman**[*]

Department of Physics, Clarkson University, Potsdam, New York 13699, USA

**ABSTRACT**

We propose the concept of autonomous self-damaging in "smart" composite materials, controlled by activation of added nanosize "damaging" capsules. Percolation-type modeling approach earlier applied to the related concept of self-healing materials, is used to investigate the behavior of the initial material's fatigue. We aim at achieving a relatively sharp drop in the material's integrity after some initial limited fatigue develops in the course of the sample's usage. Our theoretical study considers a two-dimensional lattice model and involves Monte Carlo simulations of the connectivity and conductance in the high-connectivity regime of percolation. We give several examples of local capsule-lattice and capsule-capsule activation rules and show that the desired self-damaging property can only be obtained with rather sophisticated "smart" material's response involving not just damaging but also healing capsules.

-------------------------------------------------------------------------------

[*]Corresponding author: e-mail privman@clarkson.edu, telephone +1-315-268-3891

## 1.    INTRODUCTION

There has been an increasing recent interest in "smart materials" which utilize nanoscale features to achieve useful properties, such as self-healing [1-11]. For example, in such materials development of damage and fatigue can be delayed by embedded capsules containing a microcrack-healing agent activated by a "triggering" mechanism [11]. In the present work we extend this concept to materials with self-damaging properties [12,13], as will be explained shortly, as well as to situations when both mechanisms are utilized. The first autonomous self-healing of polymer composites was realized [3] with the polymerization process initiated by the released healing agent preventing the propagation of cracks caused by mechanical stress. This finding was followed by interesting experimental [1,2,5,7,14-24] and theoretical [4,6,8-11,25-29] developments. Earlier successful approaches to self-healing required an external trigger by mechanical, physical, or chemical means [30-33].

A promising new application of smart-materials concepts could be in developing "self-damaging," also termed self-destructive, self-deteriorating, or transient materials. The first experimental realization of this property at the materials level has recently been reported for transient electronics [12,13]. Such concepts can be beneficial in many applications. They are in fact already utilized in medicine, e.g., commercially sold single-use syringes, fabrication of complex-shaped objects (self-destructive core mold materials for metal alloys [34]), as well as commercially available self-decomposing tags, labels and plastic bags. However, similar to slef-healing these earlier realizations have involved activation by external mechanical, physical (such as high temperature or laser illumination), or chemical means, rather than being internally driven by the materials' properties.

One important avenue of research has involved achieving smart-materials response at the nanoscale [4,6,7,10,11,35]. The reason for this has been that such material designs promise to allow control of the material's fatigue at the earlier stages of its development, when damage is not yet macroscopic. Material degradation [36] ultimately results in formation of voids and cracks that are macro-objects. These are initiated by the development of microscopic crazes and microcracks, the growth of which can be prevented (for self-healing) or accelerated (for self-damaging) at the nanoscale.

In this work we use a percolation modeling approach earlier utilized for self-healing [4,10,11], which considers probabilistic bond breakage in a lattice model, offering a microscopic statistical-mechanical description of the time-dependent material fatigue evolution, controlled, for instance, by releasing a "glue" substance from nanoporous capsules (in self-healing experiments [11] these were nanoporous fibers). However, we instead consider capsules causing self-damaging, as well as their combined effect with those causing self-healing. Damage formation is a multiscale phenomenon [37,38], and its modeling at different scales requires various approaches. These include continuum rate equation approach [25], as well as other continuum descriptions [29,36] involving consideration of free energies of tensile cracks or crack surface energies. More miscroscopic approaches include molecular-level modeling [29,39], discrete element methods [29], and the aforementioned "atomistic" percolation model [4,10,11] which could be suitable for understanding of certain aspects of self-healing and self-damaging at the nanoscale.

Microscopic statistical-mechanical modeling cannot always be directly related to macroscopic material's parameters, but it allows us to explore patterns of possible behaviors involving really atomic-scale effects. Here we are interested in the degree of complexity and local correlations required of the damaging (and healing) capsule activation kinetics to ensure a relatively sharp, well pronounced drop in the materials integrity after some initial fatigue developed due to the sample's usage. We use the two-dimensional (2D) lattice model earlier developed for self-healing [4], involving Monte Carlo (MC) simulations of the connectivity as a measure of the overall material's integrity. Furthermore, the actual macroscopic material's properties are typically nonlinear in the microscopic morphology measures [37], and therefore, as an example, we consider the behavior of the sample's conductivity. We point out that electrical transport properties are among the important experimentally studied macroscopic indicators of both self-damaging [12] and nanoscale self-healing [35], the latter also used as a probe of the sample's integrity. Various transport properties depend on and have been used as indicative of the material's integrity, for example, thermal conductivity [40,41], photoacoustic wave propagation [42,43], and of course electrical conductivity [35,44-47].

In addition to self-damaging, our statistical-mechanical percolation-model approach could be of interest in studies of actual networks' functionality, especially when active response/control is desired. Indeed, similar ideas involving concepts related to self-healing [48,49] and self-damaging [50] have been considered in the *sensor*-network design literature. Self-damaging concepts can be useful, for example, in designs aimed at abruptly shutting a whole computer or sensor network down if enough interconnected nodes are compromised. We point out, however, that both for materials fatigue kinetics and other applications, we are interested in self-damaging (aggressive "shutdown") after the initial, limited damage to the network or material, when the latter are still largely intact. Therefore, here we *do not consider* the critical-point behavior regime near the percolation transition, which has been of interest [51-53] in many studies of percolation models, because we are only interested in the regime of relatively high connectivity, as further commented on later. Details of our model and numerical approach are described and illustrated in Section 2. Section 3 presents results for more complicated dynamical rules. Finally, Section 4 offers a concluding discussion.

## 2. PERCOLATION MODEL OF SELF-DAMAGING

In this section we present the percolation model used in our study of features of material self-damaging with and without self-healing. Our MC simulations were carried out for a 2D lattice model similar to that which in earlier studies allowed exploration of aspects of self-healing [4,10,11]. We use a square lattice of $N^2$ sites, with periodic boundary conditions. System's degradation due to damage is described by the process of random breakage of nond connecting these sites, as detailed shortly. Some randomly positioned sites have special properties corresponding to damage-inducing or healing inclusions (capsules) in the material. The "capsule" sites can be activated and affect the kinetics of the bond connectivity in the system according to the rules defined below, as is illustrated in Figure 1.

Initially, the lattice is fully connected. We assume that during the "use" of the material, fatigue (initial damage) sets in. Here it is modeled by random breakage of bonds. Time, $t$, is measured by the number of MC sweeps through the system, i.e., $2N^2$ random bond breakage attempts are carried out per each time step. Each attempt selects a bond randomly, and, if it is intact, then it is broken with small probability $p$. Here we took the value

$$p = 0.01 \ , \tag{1}$$

comparable to earlier studies. In addition to random breakage, bonds can be broken in the vicinity of activated damaging lattice sites (capsules) or repaired near triggered healing sites. Therefore, in simulations an additional process is added in such a way that during each MC time-step, each special site is checked once on average (with random site-probing attempts). The site is activated if certain conditions typically measuring the degree of local damage buildup (detailed later for each specific modeling case) are satisfied in its predefined neighborhood, as sketched in Figure 1. The damaging or healing then affects the bonds in another predefined neighborhood, also sketched in the figure.

Generally, we then measure the material's integrity at the "microscopic" structural level by the fraction of the surviving (unbroken) bonds, $u(t)$. Unlike earlier studies, we did not assume that bonds associated with special sites have a different breakage probability. Such a choice would represent the "cost" in terms of material's integrity, i.e., some extra initial drop in the material's integrity as the trade-off of introducing the special capsules, but otherwise would not qualitatively affect the results [10].[0] As mentioned earlier, we also calculate an example of a "macroscopic" quantity, the conductivity. This was done by cutting the periodic lattice at a row of bonds, calculating the conductance, $G(t)$, using the standard approach [52] according to which a set of Kirchhoff's equations for the conducting network is solved by relaxation methods. The dimensionless conductivity is then calculated as $g(t) = G(t) / G(0)$. An example of a numerical result with simple rules (outlined shortly) is given in Figure 2.

Without any damaging or healing capsules added, the process only involving bond breakage will result in the bond connectivity of the standard bond percolation [54]. We expect the material's integrity to decay exponentially according to

$$u(t) = e^{-pt} \, , \tag{2}$$

whereas the conductance will decrease and reach zero at a finite time corresponding to the value of $u$ at the percolation transition, which in this case [54] is $\frac{1}{2}$.

Control of material's properties at the nanoscale is likely effective only in modifying damage development when defects are still really microscopic, i.e., only in the regime of the initial fatigue. Later-stage larger-scale damage development will rather be driven by macroscopic stress distributions during materials use. As a result, in the latter case our lattice statistical models, which are "cartoon" in nature and at best describe only general features of microscopic behavior, will not be applicable. Therefore, the present models can only be useful in the regime of limited damage, for instance, for material's integrity $100\% \geq u \gtrsim 80\%$, which is safely away from the percolation transition. Therefore, as emphasized earlier, here we do not consider the percolation critical behavior; our interest is only in the regime of a relatively large connectivity of the percolation cluster.

In the case illustrated in Figure 2, the "damaging" sites were activated once probed during a MC sweep if at least 3 out of the 4 bonds at that site were then already broken, but only provided that this site was not already activated earlier. Once a site was thus "activated," all the bonds connected by at least one end to any site within (and exactly at the circumference) a circle of radius 5 were broken. We found that with such a simple rule for locally uncorrelated damaging-only capsules, any significant effect can only be achieved with an unrealistically large concentration of such capsules, which here was taken 40%. The result is included in Figure 2. For the curve with the healing process also added, shown in the figure, the healing capsules' concentration was 12%, consistent with earlier studies [4], and their activation rule required at least one bond at that healing site to be broken. The healing effect on the neighboring bonds extended the same radius, 5, and each capsule could only be activated once.

Our primary objective in the present work has been to study the extent to which the self-damaging effect can be made abrupt. Ideally, as long as the material is relatively intact we would prefer to have as little self-damaging (added bond breakage) as possible. Once its integrity drops below some "tear and wear" threshold (for example, 80%) due to the onset of the initial fatigue, the self-damaging should be triggered to render the sample useless. Our simple example in Figure 2 already involves rules making the self-damaging occur only when damaged bonds are locally clustered, because we required three bonds to be earlier broken as a necessary condition for a damaging site's activation. Without such a requirement damage would set in much sooner than shown in Figure 2. Even with this requirement, the behavior of the self-damaging-only example in the figure is not exactly what we are after, because there is a loss of material's integrity for all times, even for very small fatigue. One way to repair this is to add self-healing capsules as well, which are easier broken (here, one bond breakage was required as a precondition for activation). This is seen in Figure 2, where added material's integrity degradation is avoided (as compared to the no-self-damaging/no-self-healing case) for times for which it is above approximately 80%. Here and below, we typically plot the whole curves including the percolation transition point only to show the general trend; our focus is on the initial high-connectivity behavior.

In order to have the concept of nanostructured self-damaging capsules useful in actual materials applications, the local rules of their functioning and interrelation with other active components (such as self-healing capsules) should preferably be simple, and also the concentration of the capsules of various types should be relatively small. Our simple-rule example already suggests that satisfying these expectations (simplicity of local correlations between active components and small density of these) may not be possible. In the next section we explore various modifications of the rules that yield interesting patterns of behavior.

Figure 2 also illustrates the expectation that for low damage, which is the regime of interest in the present study, various macroscopic material's properties, here, conductance, follow the material's integrity approximately linearly. Most of the results for time-dependence shown in this work, represent averages over 500 independent MC runs. Furthermore, for the regime of interest, far from the percolation threshold, we expect limited system-size dependence [4,10]. Our simulation were carried out for square lattices of sizes $N^2 = 100 \times 100$. However, we checked that finite-size effects were negligibly small for times short enough that the system was away from the percolation threshold, by running sample simulations for varying sizes up to $N = 200$. We also noted that once the percolation transition was approached, the size dependence became noticeable, but no study of this was carried out.

We point out (see also Section 3) that large fluctuations can develop in the present model if we allow "avalanche" effects. Note that the MC sweeps (MC steps) as "time" units are really arbitrary. They assume that as the material is subjected to "use," fatigue develops on the time scales of that usage. In this regime, the capsules in the simplest cases are assumed to be then activated (in experiments, literally, ruptured [7]) by the developing microscopic damage features (microcracks). However, the capsules' functioning once activated, can in principle involve their own, faster time scales of not only damaging or healing the surrounding material but also affecting the functioning of other capsules. Specifically, when their density is large enough for their domains of activity to form a percolating cluster as the damage develops, the damaging capsules can activate each other (and the healing capsules) in a chain-like effect spanning parts of the system. Our simple rules described earlier do not include such processes with independent time scales. We consider aspects of capsule-capsule activation in the next section. However, the present models are not suitable for describing large-scale damage features the dynamics of which depends on the actual macroscopic mechanical properties of the material.

## 3.     RESULTS FOR MORE COMPLICATED RULES

The model of the preceding section demonstrates that a straightwofrwad combination of self-healing and self-damaging does not lead to the desired behavior. Therefore, here we propose two more-complicated, conceptually different models. We present examples of how can local correlations involving the rules of the self-damaging and self-healing capsule "interactions" be adjusted to obtain interesting patterns of behavior, including delayed onset of a relatively sharp drop in the material's integrity and conductance. We actually tried several versions of site-site interactions, and different mechanisms and probabilities for the rules of activation of the special (damaging, healing) sites, not detailed here. The two selected cases reported here are representative of our findings. The rule considered in Section 2 had a relatively large radius, 5, for the action of activated sites, but a rather local rule for their activation. Therefore, one interesting approach consists of basing damaging capsule (site) activation on the state of its larger local neighborhood, cf. Figure 1. On the other hand, healing was sufficiently effective with the simple rule, and therefore, for our first example considered next, the activation rule of the healing capsules was not changed.

Figure 3 reports MC simulations for the model with damaging sites activated once material degradation measured by the fraction of broken bonds in the neighborhood, here taken of radius 3, exceeds a certain threshold, set at 40% for the shown results, but only if at least one of these broken bonds is at that site. For the 40% threshold count, we considered all the bonds connected to all the sites within (and exactly at) the radius 3 from the probed site. We also assumed that a damaging capsule, once activated, deactivates all the healing capsules within its activity radius, 5, and we also included some capsule-capsule instantaneous activation, the latter described in the next paragraph. In such a model, when the material is degraded to a certain degree, approximately 75% in our case, see Figure 3, curve (b), self-damaging notably accelerates the degradation process. Adding healing capsules delays the onset of this accelerated degradation process. The onset of the damage acceleration here is noticeably more abrupt than in the earlier considered example (Figure 2).

This is partly because of a more complicated, multi-bond activation rule, and also due to the added process of capsule-capsule activation. In addition to capsules being randomly probed for a possible activation during MC sweeps, we added a process with the "internal time scale" as alluded to in Section 2. This is assumed to be very fast as compared to the time scales of the external-use induced fatigue. Here we assume that only damaging capsules within a certain small radius, set at $\sqrt{5}$, of each MC-probe-activated capsule are also activated provided their own activation neighborhoods were sufficiently damaged after the latest activation, in an avalanche-like instantaneous process carried out to completion (until there are no more capsules to check in the newly damaged regions). Note that capsule-capsule activation within large enough radius can cause a system-wide avalanche effect, with sudden material's integrity collapse. This is illustrated in Figure 4 for the capsule-capsule activation radius increased to 5. However, as emphasized in Section 2, our models are not suitable for studying macroscopic damage formation. Furthermore, in order to truly establish that the "system-wide" avalanche effect here is infinite-range rather than only reaching system sizes used in our simulation (or larger, but finite), one has to carry out a finite-size scaling analysis of this phenomenon and likely gather data for much larger system sizes, which was outside the scope of the present work.

We next consider an example of rules that more aggressively rely on capsule-capsule "cross-talk." Specifically, we can condition (delay) damaging on the activation of self-healing capsules. Figure 5 presents the results for the following rules. Damaging capsules were activated provided local concentration of the used healing capsules exceeded a certain threshold, for instance, 25%, corresponding to curve (c) in Figure 5. This local region was defined by radius 3, i.e., a damaging capsule could only be triggered once 25% of healing sites in this neighborhood were already used up. (No activation at all was allowed in an unlikely case of no healing capsule present in that region.) In this model, both types of capsule can be activated if at least two bonds out of four at that site are broken. An activated damaging capsule deactivates all the healing capsules in its activity region (radius 5) thus further accelerating the degradation. This deactivation of the healing capsules meant marking them as "used," without any healing ensuing. When another damaging capsule is in the proximity to the activated damaging capsule (within radius $\sqrt{5}$ as before), it is activated immediately, but only if the "healing capsule usage" requirement is met. Curve (e) on Figure 5 represents the extreme case when damaging is not allowed, demonstrating the efficiency of the healing process alone, with these model parameters.

Generally, for models the results for which were reported in Figure 2, 3, 5, as well in in our other studies with various rule modifications, the conductance decreases approximately linearly with the material's integrity measure as long as the degree of degradation is within about 20%. For longer times and larger damage, it exhibits a percolation transition, the critical behavior at which was not of interest in our present study.

## 4. CONCLUSIONS

In summary, we studied percolation-type models with random onset of damage by bond breakage, and with various rules involving lattice sites (capsules) which can induce further damage once activated, as well as with sites which can instead cause healing. The main observation suggested by our model results is that useful patterns of self-damaging can only be obtained with rather sophisticated local rules requiring both types of capsule and involving not only the response of these capsules to the evolving material's damage but their "interactions" with each other as far as their activation is concerned.

In fact, this is not the case for self-healing, for which activation of individual capsules by simpler local rules suffices both experimentally [6,7,14,16-19,24] and in theoretical models [4,10,11,27,29] of the type considered here. The reason for this is largely in that self-healing is desirable as early in the process of the damage development as possible. On the other hand, autonomous self-damaging should preferably be minimal initially, only setting in relatively abruptly once some damage develops. The latter behavior requires more sophisticated local correlations, which might be challenging to realize in "smart" materials by using only small-fraction admixtures of active capsules. Instead, the whole material might have to be redesigned, perhaps relying on avalanche-type degradation. This indeed seems to be the case for the only experimentally reported approach to date [12]. However, as referenced earlier, similar concepts are also of interest in networks of sensors/computers [48-50], where reliance on software rather than on material's response "smart" behavior allows the use of much more complicated rules.

## ACKNOWLEDMENTS

We thank Prof. V. A. Skormin for preliminary discussions leading to the origination of the idea of the present study, and Prof. I. Sokolov for helpful suggestions. We acknowledge support of our research by the National Science Foundation under grant CBET-1066397.

# REFERENCES

[1] K. S. Toohey, N. R. Sottos, J. A. Lewis, J. S. Moore, and S. R. White, Nature Mater. **6**, 581 (2007).

[2] B. Tee, C. Wang, R. Allen, and Z. Bao, Nature Nanotech. **7**, 825 (2012).

[3] S. R. White, N. R. Sottos, P. H. Geubelle, J. S. Moore, M. R. Kessler, S. R. Sriram, E. N. Brown, and S. Viswanathan, Nature **409**, 794 (2001).

[4] A. Dementsov and V. Privman, Physica **A 385**, 543 (2007).

[5] P. Cordier, F. Tournilhac, C. Soulie-Ziakovic, and L. Leibler, Nature **451**, 977 (2008).

[6] Y. Kievsky and I. Sokolov, IEEE Trans. Nanotech. **4**, 490 (2005).

[7] J. Kirk, S. Naik, J. Moosbrugger, D. Morrison, D. Volkov, and I. Sokolov, Int. J. Fracture **159**, 101 (2009).

[8] A. C. Balazs, Materials Today **10**, 18 (2007).

[9] G. V. Kolmakov, K. Matyjaszewski, and A. C. Balazs, ACS Nano **3**, 885 (2009).

[10] A. Dementsov and V. Privman, Phys. Rev. **E 78**, 021104 (2008).

[11] V. Privman, A. Dementsov and I. Sokolov, J. Comput. Theor. Nanosci. **4**, 190 (2007).

[12] S.-W. Hwang, D.-H. Kim, H. Tao, T. Kim, S. Kim, K. J. Yu, B. Panilaitis, J.-W. Jeong, J.-K. Song, F. G. Omenetto, and J. A. Rogers, Adv. Funct. Mater. **23**, 4087 (2013).

[13] Li, R., Cheng, H., Su, Y., Hwang, S.-W., Yin, L., Tao, H., Brenckle, M. A., Kim, D.-H., Omenetto, F. G., Rogers, J. A. and Huang, Y., Adv. Funct. Mater. **23**, 3106 (2013).

[14] V. Amendola and M. Meneghetti, Nanoscale **1**, 74 (2009).

[15] E. N. Brown, N. R. Sottos, and S. R. White, Exp. Mech. **42**, 372 (2002).

[16] E. N. Brown, S. R. White, and N. R. Sottos, J. Mater. Sci. **39**, 1703 (2004).

[17] E. N. Brown, S. R. White, and N. R. Sottos, Composites Sci. Tech. **65**, 2466 (2005).

[18] E. N. Brown, S. R. White, and N. R. Sottos, J. Mater. Sci. **41**, 6266 (2006).

[19] M. R. Kessler, N. R. Sottos, and S. R. White, Composites **A 34**, 743 (2003).

[20]   P. S. Tan, M. Q. Zhang and, D. Bhattacharyya, IOP Conference Series: Materials Science and Engineering **4**, 012017 (2009).

[21]   J. W. C. Pang and I. P. Bond, Composites Sci. Tech. **65**, 1791 (2005).

[22]   A. Piermattei, S. Karthikeyan, and R. P. Sijbesma, Nature Chem. **1**, 133 (2009).

[23]   D. Shchukin and H. Möhwald, Small **3**, 926 (2007).

[24]   G. Williams, R. Trask, and I. Bond, Composites **A 38**, 1525 (2007).

[25]   E. J. Barbero, F. Greco, and P. Lonetti, Int. J. Damage Mechanics **14**, 51 (2005).

[26]   E. B. Murphy and F. Wudl, Prog. Polymer Sc. **35**, 223 (2010).

[27]   R. P. Wool, Soft Matter **4**, 400 (2008).

[28]   L. Fedrizzi, W. Fürbeth, and F. Montemor, *Self-Healing Properties of New Surface Treatments* (Maney Materials Science, Wakefield, UK, 2011).

[29]   M. Q. Zhang and M. Z. Rong, *Self-Healing Polymers and Polymer Composites* (John Wiley & Sons, Hoboken, NJ, 2011).

[30]   R. P. Wool, *Polymer Interfaces: Structure and Strength* (Hanser-Gardner Publ., Cincinnati, OH, 1995).

[31]   C. Dry, Composite Structures **35**, 263 (1996).

[32]   C. B. Lin, S. Lee, and K. S. Liu, Polymer Eng. Sci. **30**, 1399 (1990).

[33]   J. Raghavan and R. P. Wool, J. Appl. Polym. Sci. **71**, 775 (1999).

[34]   K. S. Mazdiyasni and R. R. Wills, *Self-destructive core mold materials for metal alloys*, Patent: US4043381 (1977).

[35]   E. T. Thostenson and T.-W. Chou, Adv. Mater. **18**, 2837 (2006).

[36]   R. K. Bregg, *Frontal Polymer Research* (Nova Science Publ., NY, 2006).

[37]   G. W. Milton, *The Theory of Composites* (Cambridge University Press, Cambridge, UK, 2002).

[38]   K. B. Broberg, *Cracks and Fracture* (Cambridge University Press, Cambridge, UK, 1999).

[39]    S. Maiti, C. Shankar, P. H. Geubelle, and J. Kieffer, J. Eng. Mater. Technol. **128**, 595 (2006).

[40]    I. Sevostianov, Int. J. Eng. Sci. **44**, 513 (2006).

[41]    I. Sevostianov and M. Kachanov, Adv. Appl. Mechanics **42**, 69 (2009).

[42].   M. Navarrete, M. Villagrán-Munizb, L. Poncec and T. Flores, Opt. Lasers Eng. **40**, 5 (2003).

[43]    A. S. Chekanov, M. H. Hong, T. S. Low and Y. F. Lu, IEEE Trans. Magn. **33**, 2863 (1997).

[44]    M. Kupke, K. Schulte and R. Schüler, Composites Sci. Tech. **61**, 837 (2001).

[45]    R. Schueler, S. P. Joshi, and K. Schulte, Composites Sci. Tech. **61**, 921 (2001).

[46]    K. Schulte and C. Baron, Composites Sci. Tech. **36**, 63 (1989).

[47]    I. Weber and P. Schwartz, Composites Sci. Tech. **61**, 849 (2001).

[48]    M. Coles, D. Azzi, and B. Haynes, Sensor Review **28**, 326 (2008).

[49]    Y. Qu and S. Georgakopoulos, Wireless Sensor Network **4**, 257 (2012).

[50]    A. R. Chowdhury, S. Tripathy, and S. Nandi, in: Proc. COMPSWARE 2007, *2nd International Conference on Communication Systems Software and Middleware*, pp. 1-5, DOI: 10.1109/COMSWA.2007.382600 (2007).

[51]    H. E. Stanley, *Introduction to Phase Transitions and Critical Phenomena* (Oxford University Press, Oxford, UK, 1993).

[52]    S. Kirkpatrick, Rev. Mod. Phys. **45**, 574 (1973).

[53]    C. D. Lorenz and R. M. Ziff, Phys. Rev. **E 57**, 230 (1998).

[54]    D. Stauffer and A. Aharony, *Introduction to Percolation Theory* (CRC Press, Boca Raton, FL, 1994).

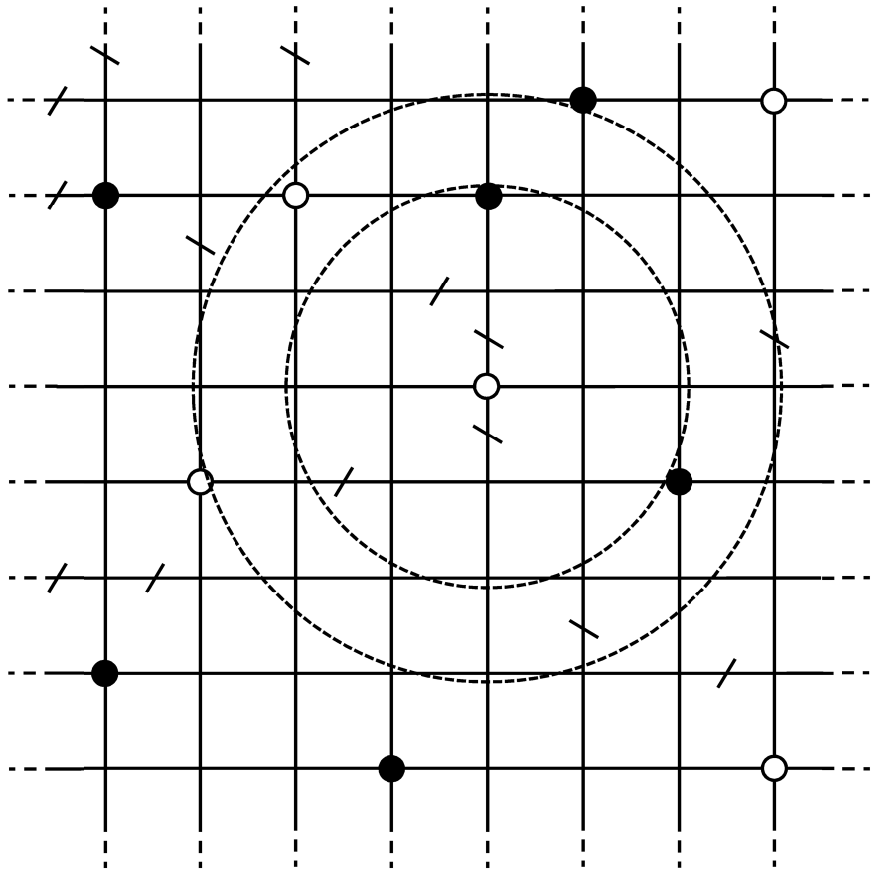

**Figure 1.** An example of a square lattice with periodic boundary conditions, with a fraction of sites with special properties, randomly positioned, representing damaging (small open circles) or healing (small filled circles) capsules. These active sites have two neighborhoods associated with them, here schematically represented by the two large dotted circles, one of the range of their effect if activated, and another used to decide on their activation. Specific rules are detailed in the text. The broken bonds (shown crossed) indicate the presence of the material's "damage."

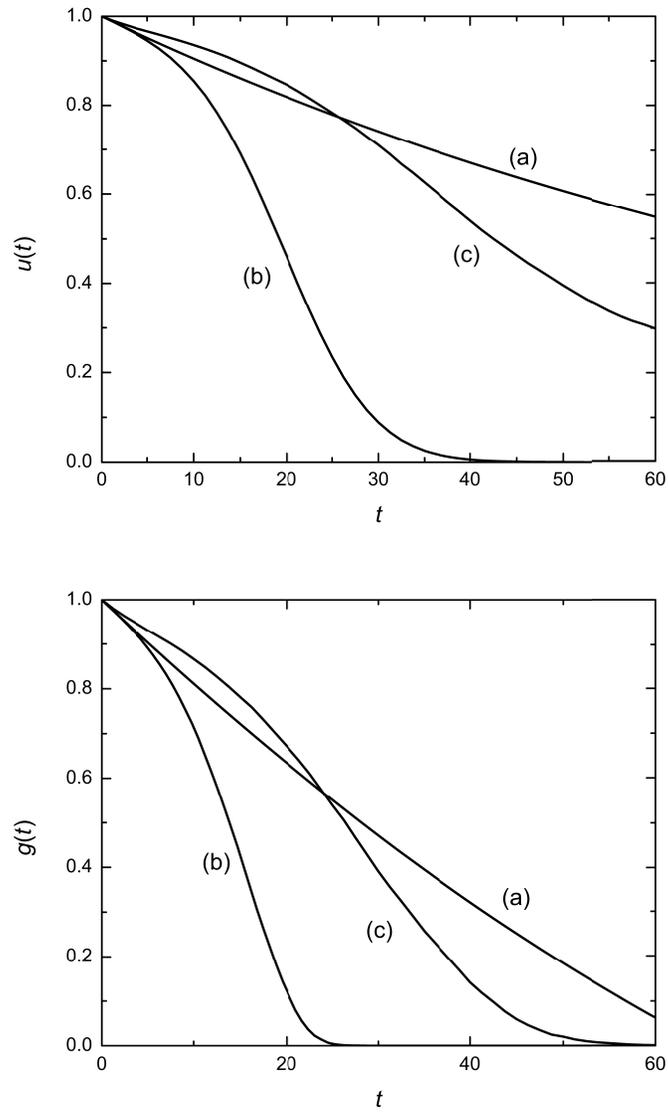

**Figure 2.** Illustration of the behavior of the material's integrity measure (top panel) and conductance (bottom panel) for a model with relatively simple rules: (a) only random bond breakage, corresponding to the standard bond percolation; (b) with damaging capsules added; and (c) with healing capsules also present, in addition to damaging.

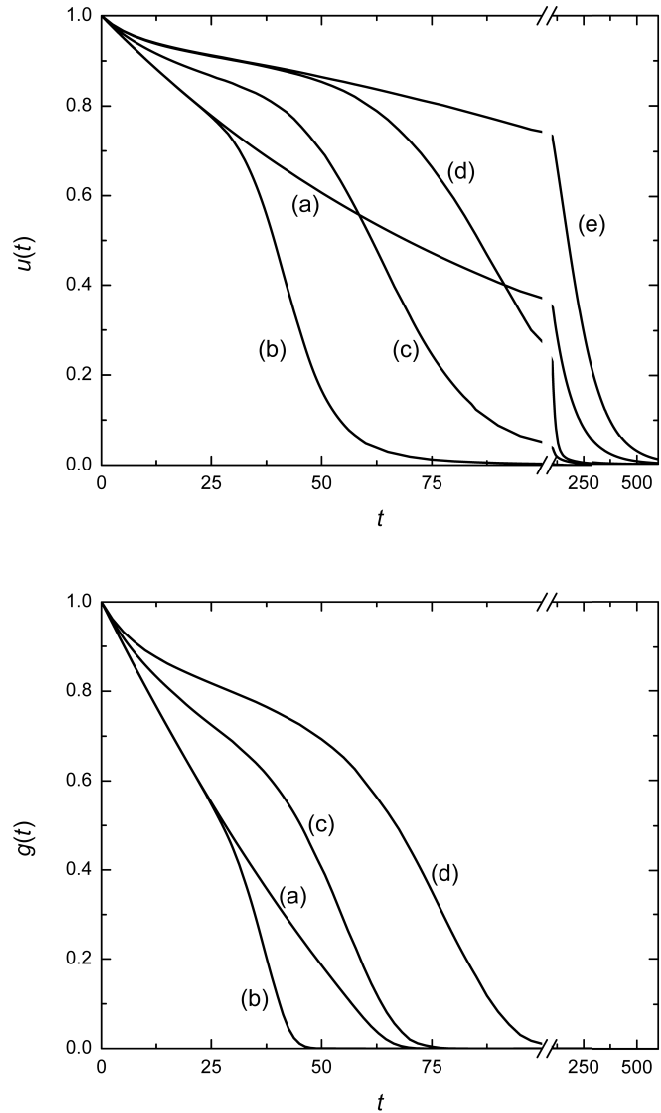

**Figure 3.** Demonstration of the approach when local material's degradation, measured by the fraction of broken bonds, activates damaging, with added damaging capsule activation and healing capsule deactivation by damaging capsules. Detailed rules for the damaging and healing processes are given in the Section 3. (a) The case with no damaging or healing processes, shown for reference. (b) Damaging only, with the initial concentration of the damaging sites of 10%. (c) The same, with 5% concentration of added healing sites. (d) Damaging, with 10% concentration of added healing sites. (e) Healing only, 10%, without damaging sites present. The conductance for this case is not shown because it follows the material's integrity approximately linearly, without any interesting features for time scales up to $t \approx 100$.

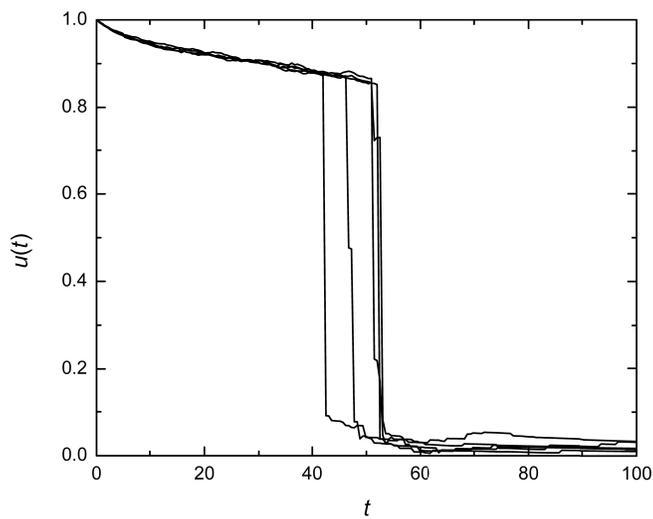

**Figure 4.** Illustration of the model similar to that in Figure 3, with 10% of healing capsules, but with the damaging capsule avalanche capsule-capsule activation radius increased to 5, as described in Section 3. Shown are five different MC runs, rather than the average over many runs as in all the other figures. The system illustrates large fluctuations due to sudden bursts of the spread of damage over substantial fractions of its size.

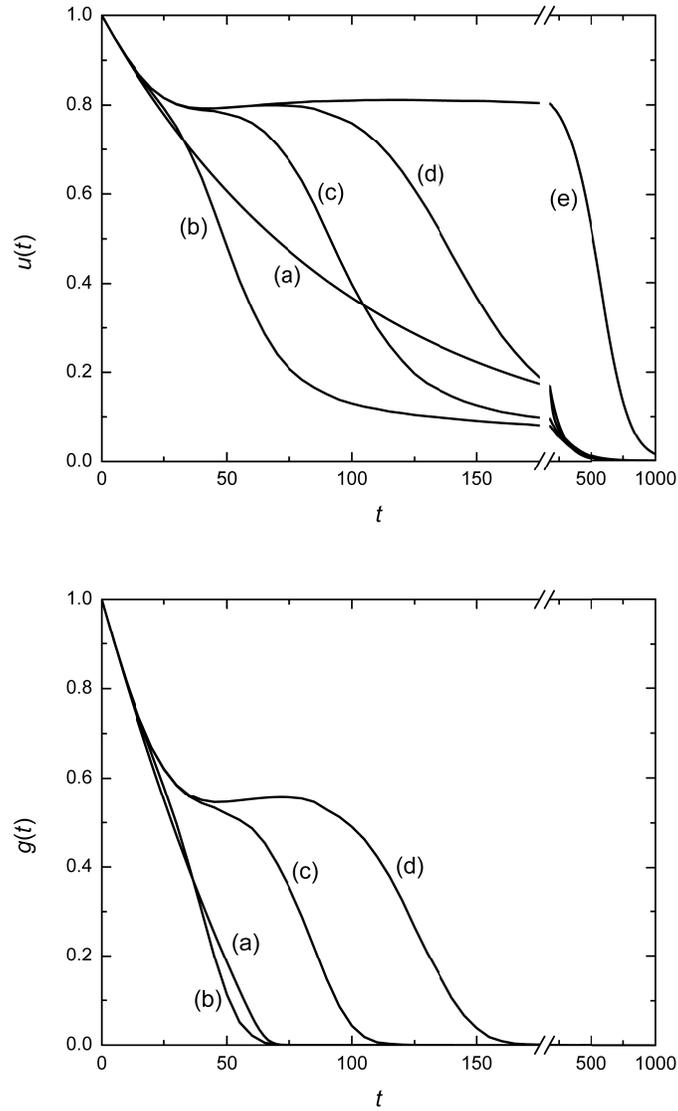

**Figure 5.** Results for a model with damaging conditioned of the activation of healing capsules, as detailed in the text. (a) The case with no damaging or healing. For curves (b), (c), and (d), the initial concentration of the damaging capsules was 10% and of healing capsules 20%. (b) The damage activation threshold was 10% of the healing capsules used up in a local neighborhood, see text; (c) 25% used up; (d) 50% used up. (e) The healing process only, with only healing sites present (20% initial concentration). The conductance for this case is not shown because it follows the material's integrity approximately linearly, without any interesting features, for time scales up to $t \approx 200$.